\documentclass[superscriptaddress,showpacs,amssymb,10pt,reprint,aps,prd,longbibliography,nofootinbib,floatfix]{revtex4-2}
\usepackage{ulem}
\usepackage{graphicx,epsfig,amssymb} 
\usepackage{amsmath,amsfonts, times}
\usepackage{bm} 
\usepackage{epstopdf}
\usepackage[linktocpage,colorlinks=true,urlcolor=blue,linkcolor=blue,citecolor=blue]{hyperref}
\usepackage[caption=false]{subfig}
\usepackage[usenames]{color}     
\usepackage{natbib}
\usepackage{soul}
\usepackage[utf8x]{inputenc}
\usepackage[usenames]{color}
\usepackage{float}
\usepackage{nccmath}
\definecolor{darkgreen}{rgb}{0,0.5,0}

\def\p{\partial}
\newcommand{\sss}{\scriptscriptstyle}
\def\Q{\text{\textit{Q}}}
\def\KN{\text{KN}}
\def\KS{\text{KS}}

\begin{document}
	\title{Scattering by stringy black holes}
		
	\author{S\'ergio V.~M.~C.~B.~Xavier}
			\email{sergio.xavier@icen.ufpa.br}
	\affiliation{Programa de P\'os-Gradua\c{c}\~{a}o em F\'{\i}sica, Universidade 
		Federal do Par\'a, 66075-110, Bel\'em, Par\'a, Brazil.}
	\affiliation{Departamento de Matem\'atica da Universidade de Aveiro and Centre for Research and Development in Mathematics and Applications (CIDMA), Campus de Santiago, 3810-183 Aveiro, Portugal.}

	\author{Carolina L. Benone}
	\email{benone@ufpa.br} 
	\affiliation{Campus Salinópolis,
		Universidade Federal do Par\'a, 68721-000, Salinópolis, Par\'a, Brasil}
	
	\author{Luiz~C.~S.~Leite}
	\email{luiz.leite@ifpa.edu.br}
	\affiliation{Campus Altamira, Instituto Federal do Pará, 68377-630, Altamira, Pará, Brazil}
	
	\author{Lu\'is C.~B.~Crispino}
		\email{crispino@ufpa.br}
	\affiliation{Programa de P\'os-Gradua\c{c}\~{a}o em F\'{\i}sica, Universidade 
		Federal do Par\'a, 66075-110, Bel\'em, Par\'a, Brazil.}
	\affiliation{Departamento de Matem\'atica da Universidade de Aveiro and Centre for Research and Development in Mathematics and Applications (CIDMA), Campus de Santiago, 3810-183 Aveiro, Portugal.}

	\begin{abstract}
	We study the scattering of axially incident massless scalar waves by a charged and rotating black hole solution from heterotic string theory called Kerr--Sen black hole. We compute the scattering cross section using the partial wave approach, for arbitrary incident wavelengths. We compare our results with those of the general relativistic version of a charged and rotating black hole, namely the Kerr–Newman black hole. We present a selection of numerical results showing that these compact objects have similar scattering properties.
\end{abstract}
	
	\date{\today}
	
	\maketitle

	\section{Introduction}\label{sec:int}
 The scattering of particles and fields is a relevant theme in physics.  This phenomenon can help us to investigate the constituents of the Universe and the fundamental interactions in nature. Scattering plays a fundamental role in particle accelerators, such as the Large Hadron Collider (LHC), which is going on its third run \cite{LHC:2022}.
		
		 Studying the scattering of fields by black holes (BHs) allows us to better understand the properties of spacetime itself. 
In the realm of BH physics, the interaction that governs the wave scattering by these compact objects is primarily gravitational. 
However, in some BH solutions, non-gravitational fields can play a non-trivial role in these events. 
		
Several works have considered the scattering by BHs described by General Relativity (GR) theory.  
The kickoff on this field is due to Matzner~\cite{Matzner:1968}, who studied the scattering and absorption of massless scalar waves by a Schwarzschild BH. Other works completed the characterization of the wave scattering and absorption by neutral GR BHs (cf. Refs~\cite{Vishveshwara:1970, Sanchez:1978, Anderson:1995, Jung_Park:2004,Crispino_etal:2009b,LDC:2017} and references therein), including the glory effect~\cite{Matzner_etal:1985} -- the presence of a bright spot or ring around the scattering object due to wave interference~\cite{Leite_etal:PRD2019}. 

		The electrically charged version of the Schwarzschild BH and its rotating generalization were also targets of investigation in scattering and absorption phenomena, as can be seen in Refs.~\cite{Handler_Matzner: 1980,Glampedakis_Anderson:2001,Crispino_etal:2009,CHM:2010,OCH:2011,CDHO:2014,CDHO:2015,Benone:2018rtj,BC:2019,Leite_etal:PRD2019,Leite_etal:2019}. These studies revealed the influence of angular momentum and electrical charge on the scattering spectrum of BHs.
		
		One can explore the scattering of fields by BH solutions not described by GR in order to seek possible signatures of different theories of gravity. In the last few years, a BH solution known as Kerr-Sen (KS) BH has been attracting attention \cite{Sen:PRL1992}. It is a charged and rotating BH solution of the low energy limit of heterotic string theory, a candidate for quantum gravity theory. Many aspects of the KS spacetime have already been investigated, such as the shadow \cite{Younsi_etal:2016,Xavier_etal:2020}, superradiant instability \cite{Bernard:2017}, clouds \cite{Bernard:2016}, hidden symmetries \cite{Hioki_Miyamoto:2008}, cosmic censorship \cite{Gwak:2017,Siahaan:2016} and merger estimates \cite{Siahaan:2020}. Also, some constraint on the dilaton charge was found by analyzes of X-ray spectroscopy \cite{Tripathi_etal:2021}.
		
		Only recently, the absorption of a massless scalar wave by a KS BH (KSBH) was investigated \cite{Xavier_etal:2021}. In Ref.~\cite{Huang_Zhang:2020}, the authors studied the scattering of fields by the non-rotating version of KSBH. Nonetheless, to the extent of our knowledge, a careful investigation of wave scattering by KSBH is still lacking in the literature.
		
We explore the scattering of massless scalar waves by such charged and rotating stringy BH solution, namely, the KSBH. We numerically compute the total scattering cross section using the partial wave approach. We compare our numerical findings with the classical differential cross section and an approximation for the glory scattering. We contrast the results of KSBH with a correspondent solution of a charged and rotating BH in GR theory, i.e., the Kerr-Newman (KN) spacetime~\cite{Leite_etal:2019}.

		The remainder of this paper is organized as follows. In Sec.~\ref{sec:1}, we introduce the charged rotating black hole spacetimes analyzed in this work, namely, KSBH and KN BH (KNBH). In Sec.~\ref{sec:2}, we discuss the massless scalar field dynamics in the KS spacetime. We show the partial wave approach and the expression for the scattering cross section. Additionally, we exhibit a brief review of the null geodesic scattering problem and the glory effect. In Sec.~\ref{sec:3}, we present a selection of our numerical results of the scattering cross section for different values of the black hole charge and spin. We also contrast our results with those obtained in the KN case. We present our final remarks in Sec.~\ref{sec:remarks}. Throughout the paper we use natural units ($c=G=\hbar=1$) and the metric signature ($+,-,-,-$).

\section{Charged rotating black holes} \label{sec:1}

The leading BH solution explored in this work is a solution to the low-energy limit of the heterotic string theory. Additionally to the metric field $g_{\mu\nu}$, the action of this theory is formed by a dilaton field $\tilde{\Phi}$ and a third-rank tensor $H_{k\mu\nu}$ associated with a Kalb-Ramond field $B_{\mu\nu}$. In the Einstein-frame, the action reads~\cite{Sen:PRL1992,Delgado_Herdeiro_Radu:2016}:
	\begin{equation}
		\begin{split}
			S = \int d^{4}x\sqrt{-g}\left(R-e^{-2\tilde{\Phi}}F_{\mu\nu}F^{\mu\nu}-2\p_{\mu}\tilde{\Phi}\p^{\mu}\tilde{\Phi}\right.\\
			\left. -\frac{1}{12}e^{-4\tilde{\Phi}}H_{k\mu\nu}H^{k\mu\nu}\right),
		\end{split}
	\label{actionST}
	\end{equation}
where $R$ is the Ricci scalar, $F_{\mu\nu}=\p_{\mu}A_\nu-\p_{\nu}A_\mu$ is the Maxwell field strength with $A_\mu$ being the potential, and the tensor $H_{k\mu\nu}$ is defined as
	\begin{align}
		H_{k\mu\nu}\equiv &\p_{k}B_{\mu\nu}+\p_{\nu}B_{k\mu}+\p_{\mu}B_{\nu k}\nonumber\\
		&-2\left(A_{k}F_{\mu\nu}+A_{\nu}F_{k\mu}+A_{\mu}F_{\nu k}\right).
	\end{align}

	In Boyer-Lindquist coordinates $(t,r,\theta, \varphi)$, the line element of the stationary and axisymmetric KS solution takes the following form:
	\begin{align}
		ds^{2}=&\left(1-\frac{2 M r}{\rho^{2}}\right)dt^{2}-\rho^{2}\left(\frac{dr^{2}}{\Delta}+d\theta^{2}\right)+\nonumber\\
		&\frac{4Mra\sin^{2}\theta}{\rho^{2}}dtd\varphi-\left(\xi+\frac{2 M r a^{2}\sin^{2}\theta}{\rho^{2}}\right)\sin^{2}\theta\,d\varphi^{2},
		\label{ds2KS}
	\end{align}
where
	\begin{align}
		&d\equiv \frac{q^{2}_{\sss \KS}}{2M},\\
		&\Delta\equiv r\left(r+2d\right)-2 M r+a^{2},\\
		&\rho^{2}\equiv r\left(r+2d\right)+a^{2}\cos^{2}\theta,\\
		&\xi\equiv r\left(r+2d\right)+a^{2}.
	\end{align}
The parameters $(M, q_{\sss \KS}, a)$ are, respectively, the mass, the electric charge and the angular momentum per unit mass of KSBH.	
	
The solutions for the dilaton, electromagnetic and axion fields of this spacetime are:
	\begin{align}
		&\tilde{\Phi}=-\frac{1}{2}\ln\frac{\rho^{2}}{r^2+a^2\cos^2\theta},\\
		&A_\mu dx^{\mu}=\frac{q_{\sss \KS}}{\rho^2}r(dt-a\sin^{2}\theta d\varphi),\\
		&B_{t\varphi}=\frac{q^{2}_{\sss \KS}ra\sin^{2}\theta}{2M\rho^2}.
	\end{align}
	
 The event horizon of KSBH is localized at
	\begin{equation}
		r_{h}^{\sss \text{KS}} \equiv M-d+\sqrt{\left(M-d\right)^2-a^2}. 
		\label{rh_ks}
	\end{equation}

We compare the scattering results of KSBH with a charged and rotating BH solution obtained within GR theory, namely, KNBH. In Boyer--Lindquist coordinates, the KNBH line element reads:
		\begin{align}
	ds^{2}=&\left(1-\frac{2 M r - q_{\sss \KN}^{2}}{\rho^{2}_{\KN}}\right)dt^{2}-\rho^{2}_{\KN}\left(\frac{dr^{2}}{\Delta_{\KN}}+d\theta^{2}\right)\nonumber\\
	&+\frac{4 M a r\sin^{2}\theta-2a q_{\sss \KN}^{2}\sin^{2}\theta}{\rho^{2}_{\KN}}dtd\varphi\nonumber\\
	&-\frac{\left[(r^{2}+a^{2})^{2}-\Delta_{\KN} a^{2}\sin^{2}\theta\right]\sin^{2}\theta}{\rho^{2}_{\KN}}d\varphi^{2},
	\label{ds2KN}
	\end{align}	
	where
	\begin{align}
	&\Delta_{\text{KN}}\equiv r^{2}-2 M r+a^{2}+q_{\sss \KN}^{2},\\
	&\rho^{2}_{\text{KN}}\equiv r^{2}+a^{2}\cos^{2}\theta.
	\end{align}

In the KN spacetime, the electromagnetic vector potential is given by
	\begin{equation}
	A_{\mu}dx^{\mu}=\frac{q_{\sss \KN}}{\rho^{2}_{\sss \KN}}r(dt-a \sin^{2}\theta d\varphi). 
	\label{A_KN}
	\end{equation}		
	
The KNBH presents an event horizon at:
	\begin{equation}
	r_{h}^{\sss \KN} \equiv M +\sqrt{M^{2}-a^{2}-q_{\sss \KN}^{2}}.
	\label{rh_kn}
	\end{equation}

\begin{figure}
	\includegraphics[width=8cm]{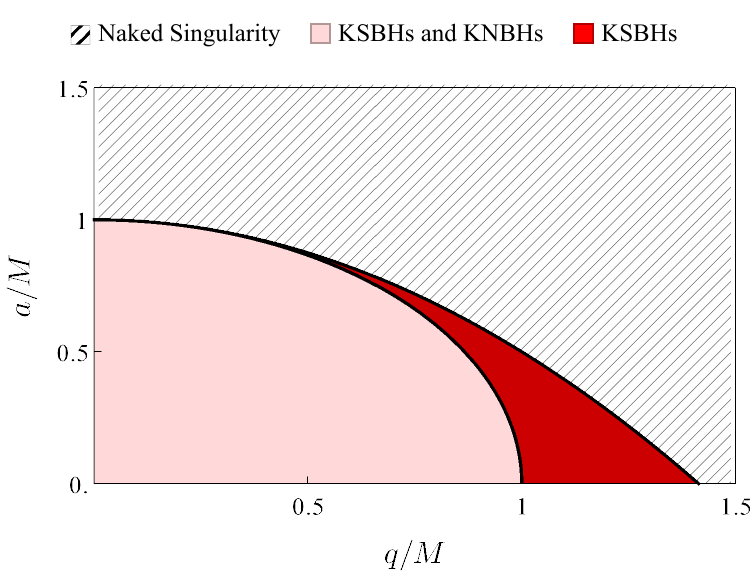}
	\caption{\label{ChargeKNvsKS} Domain of existence of BH solutions of KN and KS spacetimes.  The solid black lines correspond to extremal BHs. Notice that the KS solution permits a broader range of values for the electric charge associated with BHs.
}
	\end{figure}

Both KS and KN spacetimes have similar geometrical aspects, but also present some peculiar distinctive features \cite{Burinskii:1995}. For instance, from Eq.~\eqref{rh_ks} and Eq.~\eqref{rh_kn}, it is possible to see that the KS solution allows the existence of BHs with a wider range of values of electric charge than the KN solution, as illustrated in Fig.~\ref{ChargeKNvsKS}.

In order to help in the comparison between the scattering results of KSBHs and KNBHs, we introduce the normalized charge $\Q$ as
	\begin{equation}
	\Q_{(i)}=q_{(i)}/q^{ext}_{(i)},
\end{equation}	
 where $q^{ext}_{(i)}$ is the value of the charge of an extreme BH, and the subscript $(i)=\KS,\KN$ indicates the corresponding spacetime.
	\color{black}

\section{Scattering cross section} \label{sec:2}

\subsection{Planar-wave scattering}
We are interested in studying the scattering of massless and chargeless scalar plane waves $\Psi (x^{\mu})$ by charged and rotating BHs. These scalar fields satisfy the Klein-Gordon equation, which, in its covariant form, can be written as
	\begin{equation}
		\frac{1}{\sqrt{-g}}\p_{\mu}(\sqrt{-g}g^{\mu\nu}\p_{\nu}\Psi)=0,
		\label{KGeq}
	\end{equation}
	where $g$ is the determinant of the KS metric, whose contravariant components are $g^{\mu\nu}$.
	
Since we are concerned with monochromatic plane waves, we can use the following ansatz to solve Eq.~\eqref{KGeq}:
		\begin{equation}
			\Psi(x^\mu)=\sum_{l=0}^{+\infty}\sum_{m=-l}^{+l}\frac{U_{\omega l m}(r)}{\sqrt{r(r+2d)+a^{2}}}S_{\omega l m}(\theta)e^{i(m\varphi-\omega t)}.
			\label{Psi}
		\end{equation}
		\normalsize
The functions $S_{\omega l m}(\theta)$  are the oblate spheroidal harmonics obtained by the resolution of the equation \cite{Abramovitz}
	\begin{align}
		&\left(\frac{d^{2}}{d\theta^{2}}+\cot\theta\frac{d}{d\theta}\right)S_{\omega lm}\nonumber\\
		&+\left(\lambda_{lm}+a^{2}\omega^{2}\cos^{2}\theta-\frac{m^{2}}{\sin^{2}\theta}\right)S_{\omega lm}=0,
	\end{align}
where $\lambda_{lm}$ are the eigenvalues of the spheroidal harmonics. 

The radial solution $U_{\omega lm}(r)$ satisfies the following equation:
	\begin{equation}
		\left(\frac{d^{2}}{dr^{2}_{\star}}+V_{\omega lm}\right)U_{\omega lm}(r_{\star})=0,
		\label{RWeq}
	\end{equation}
	where $r_{\star}$ is the tortoise coordinate, defined as
	\begin{equation}
		r_{\star}\equiv \int dr\left(\frac{r(r+2d)+a^{2}}{\Delta}\right).
	\end{equation}
The function $V_{\omega lm}$ is given by:
	\begin{widetext}
	\begin{align}
		V_{\omega lm}=&-\frac{\Delta  \left(a^2 \left(-d^2-2 d M+2 d r-4 M r+r^2\right)+a^4+r \left(d^2 (2 M-r)-2 d^3+2 d M r+2 M r^2\right)\right)}{\left(a^2+2 d r+r^2\right)^4}\nonumber\\
		&-(\lambda _{lm} +a^2 \omega^2 -2 m a\omega)\frac{\Delta }{\left(a^2+2 d r+r^2\right)^2}+\frac{\left(a^2 \omega -a m+2 d r \omega +r^2 \omega \right)^2}{\left(a^2+2 d r+r^2\right)^2}.
		\label{V_KS}
	\end{align}\\
\end{widetext}
			
	The appropriate boundary conditions to our scattering problem, to solve Eq.~\eqref{RWeq},  are dictated by the so-called {\it in} modes. They represent incoming waves from the past null infinity, part of which is transmitted to the event horizon while the remaining portion is reflected to the future null infinity. These $U_{\omega lm}$ modes are such that:
	\begin{equation}
		U_{\omega lm}(r_{\star})~\begin{cases}
		\mathcal{I}_{\omega lm}U_{I}+\mathcal{R}_{\omega lm}U^{*}_{I} \hspace{0.3cm} (r_{\star}/M\rightarrow +\infty),\\
		\mathcal{T}_{\omega lm}U_{T} \hspace{1.9cm} (r_{\star}/M\rightarrow -\infty),
		\end{cases}
	\end{equation}
with the symbol $*$ denoting complex conjugation. The terms $U_{I}$ and $U_{T}$ can be expanded as:
	\begin{align}
		&U_{I}=e^{-i\omega r_{\star}}\sum_{j=0}^{N}\frac{h_{j}}{r^{j}},\\
		&U_{T}=e^{-i(\omega-m\Omega_{H})r_{\star}}\sum_{j=0}^{N}g_{j}(r-r_{h}^{\sss \KS})^{j},
	\end{align}
where 
	\begin{equation}
		\Omega_{H}=\frac{a}{2 Mr_{h}^{\sss \KS}}
	\end{equation}
is the angular velocity of the event horizon.
The  coefficients $h_{j}$ and $g_{j}$ are constants found by solving Eq.~\eqref{RWeq}.

The main goal of a scattering problem is to obtain the differential scattering cross section, given in terms of the scattering amplitude $f(\theta,\varphi)$, namely \cite{Newton:1982}:
	\begin{equation}
		\frac{d\sigma}{d\Omega}= |f(\theta,\varphi)|^2.
		\label{partialabs}
	\end{equation}

For the on-axis case (when the wave is incident along the axis of rotation of the BH), one can write the scattering amplitude as
	\begin{equation}\label{scatt_amplitude}
	f(\theta) = \frac{2\pi}{i\omega}\sum_{l=0}^{\infty}S_{\omega l 0}(0)S_{\omega l 0}(\theta)\left[(-1)^{l+1}\mathcal{R}_{\omega l 0}-1\right],
	\end{equation}
with $\theta$ being the scattering angle. 
Note that, due to the axial symmetry of this problem, the result is independent of the azimuthal number $m$.

	To compare the scattering cross section between the two different spacetimes, we define the ratio between the scattering cross section of KSBHs and KNBHs as
	\begin{equation}
	\Sigma=\frac{(d\sigma/d\Omega)_\KN}{(d\sigma/d\Omega)_\KS}.\label{eq:ratio}
	\end{equation}
	
\subsection{Geodesic scattering}

Null geodesics in KS spacetime are described by the following equations:
\begin{fleqn}
\begin{align}
&\rho^{2}_{{\sss \KS}}\dot{t}=\frac{\xi^{2}-2 M r a \mathcal{L}}{\Delta}-a^{2}\sin^{2}\theta,
		\label{tcarter}\\
		&\rho^{2}_{{\sss \KS}}\dot{\varphi}=\frac{\mathcal{L}- a\sin^2\theta}{\sin^{2}\theta}+\frac{a\left(\xi-a\mathcal{L}\right)}{\Delta},
		\label{phicarter}\\		
		&\rho^{4}_{{\sss \KS}}\dot{r}^2=\left(a\mathcal{L}-\xi\right)^{2}-\Delta\left[\left(\mathcal{L}-a\right)^{2}+\mathcal{K}\right],
		\label{rcarter}\\
		&\rho^{4}_{{\sss \KS}}\dot{\theta}^2=\mathcal{K}-\cos^{2}\theta\left[\frac{\mathcal{L}^{2}}{\sin^{2}\theta}-a^{2}\right],
		\label{thetacarter}
	\end{align}
	\end{fleqn}
where $\mathcal{L}\equiv L/E$ is the axial component of the angular momentum per unity energy of the massless particle, 
 as measured by an observer at infinity; and  $\mathcal{K}\equiv K/E^2$ is a separation constant per unity energy, equivalent to the one introduced by Carter \cite{Carter:1968}.

For the on-axis case $\mathcal{L}=0$ and the impact parameter (the orthogonal distance between the path of the particle and the center of the BH \cite{Goldstein:1950}) is:
	\begin{equation}
	b = \sqrt{\mathcal{K}+a^2}.
	\end{equation}
	The orbital equation can be written as:
	\begin{equation}\label{orbital_eq}
\left(\frac{dr}{d\theta}\right)^2=\frac{\left[r\left(r+2d\right)+a^2\right]^2-b^2\Delta_{KS}}{b^2-a^2\sin^2\theta}.
	\end{equation}
	One can rewrite  Eq.~\eqref{orbital_eq} defining $u\equiv1/r$, namely
	\begin{equation}
\left(\frac{du}{d\theta}\right)^2=u^4\left[\frac{(u^{-2}+2d/u+a^2)^2-b^2\Delta_u}{b^2-a^2\sin^2\theta}\right],
	\end{equation}
	where $\Delta_u\equiv \left(u^{-1}+2d\right)/u-2 M/u+a^2$.
	
	The deflection angle is defined as \cite{Collins_etal:1973}
	\begin{equation}
	\Theta(b)\equiv 2\beta(b)-\pi,
	\end{equation}
	with
		\begin{equation}
		\beta(b)=\int_0^{u_0}\frac{1}{u^2}\sqrt{\frac{b^2-a^2\sin^2\theta}{(u^{-2}+2d/u+a^2)^2-b^2\Delta_u}},
		\end{equation}
	and $u_0$ is associated with the radius $r_0$ of the closest approach for a null geodesic. 
	The relation between the scattering angle and the deflection angle is expressed by:
		\begin{equation}
		\Theta=\pm\theta-2N\pi,\hspace{0.3cm}\text{with}\hspace{0.2cm}N=0,1,2,\cdots
		\end{equation}
	
	Considering an incident beam of particles, the geodesic differential scattering cross section is given by:
	\begin{equation}
	\frac{d\sigma}{d\Omega}=\frac{1}{\sin\theta}\sum_j b_j(\theta)\left|\frac{db_j}{d\theta}\right|.
	\label{geo_scatt_CS}
	\end{equation}
	The function \(b(\theta)\) displays a multi-valued nature with respect to \(\theta\), given that particles can undergo multiple revolutions around the BH. Consequently, when evaluating the sum, it becomes essential to consider distinct \(b_j\) values that correspond to the same scattering angle \(\theta\). These different \(b_j\) values collectively contribute to the cross section \eqref{geo_scatt_CS}.

\subsection{Glory scattering}

\begin{figure}
	\includegraphics[width=8cm]{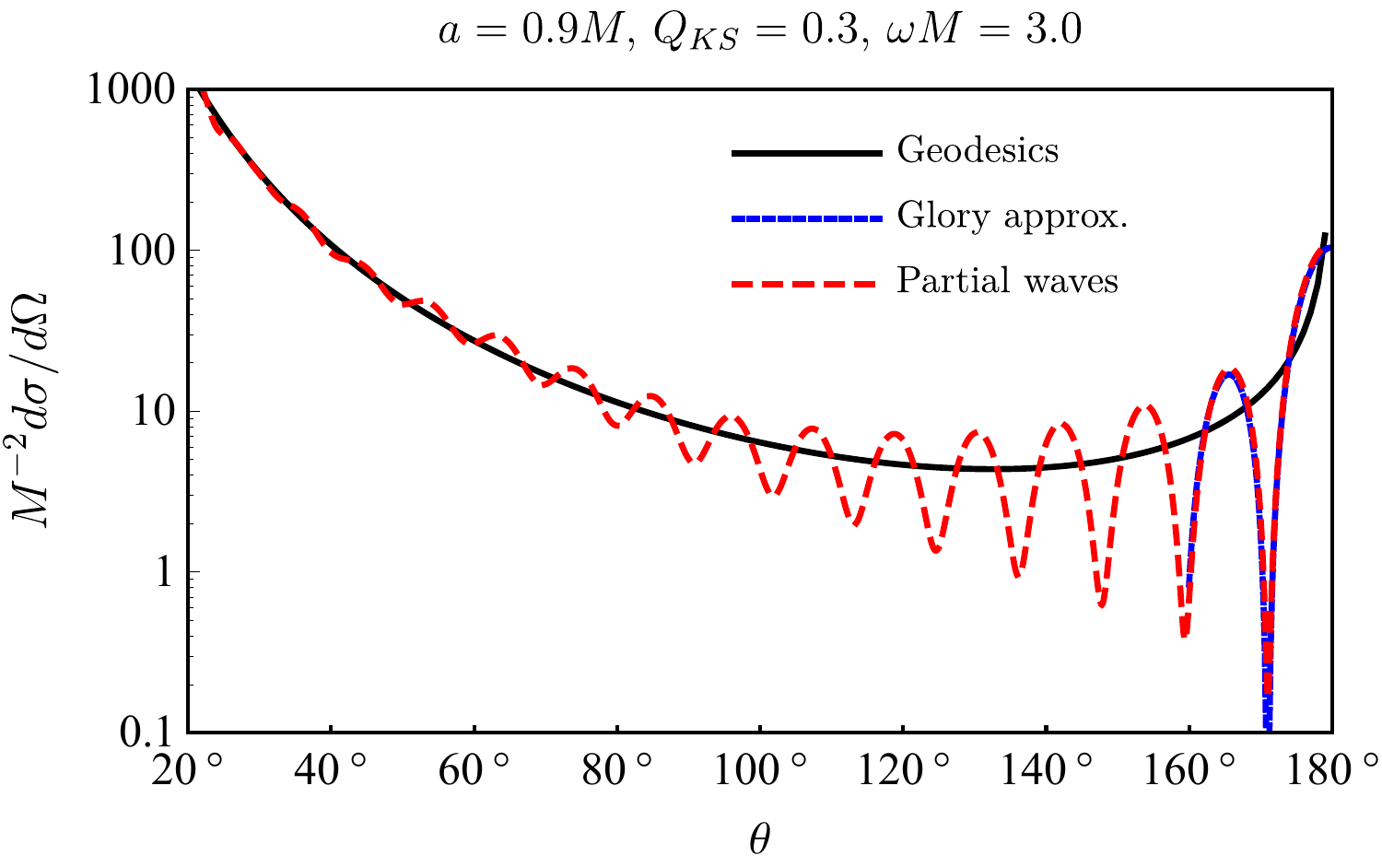}
	\caption{\label{cls_geo_nscs}
	Comparison between the scalar and classical KS scattering cross section for $a/M=0.9$, $\Q_{KS}=0.3$, and $M\omega=3.0$. The glory approximation is also exhibited. 
}
	\end{figure}

A bright spot or halo in the backward direction can characterize the scattering at large angles. This arises due to the interference between waves traveling in opposite senses around the BH. Near the backward direction, the scalar scattering cross section can assume the following analytic form~\cite{Matzner_etal:1985}
	\begin{equation}
	\frac{d\sigma}{d\Omega}\bigg|_{\theta\approx\pi}=\mathcal{A}J_0^2(\omega b_g\sin\theta),
	\end{equation}
	where $J_0$ is the Bessel function of the first kind of order $0$ and
	\begin{equation}
	\mathcal{A}\equiv2\pi\omega b_g\left|\frac{db}{d\theta}\right|_{\theta=\pi},
	\end{equation}
with $b_g\equiv b(\pi)$ being the glory impact parameter.

In Fig.~\ref{cls_geo_nscs}, we compare the scalar and the geodesic scattering cross sections of a KSBH for $a/M = 0.9$, $\Q_{\KS} = 0.3,$ and $M\omega = 3.0$. We also present the glory approximation. The long-range nature of the gravitational field results in a divergence of the scattering cross section for small angles. Furthermore, we notice a good agreement between the scalar scattering cross section and the semiclassical glory approximation near the backward direction.

\section{Results}\label{sec:3}

\begin{figure*}
\includegraphics[width=8cm]{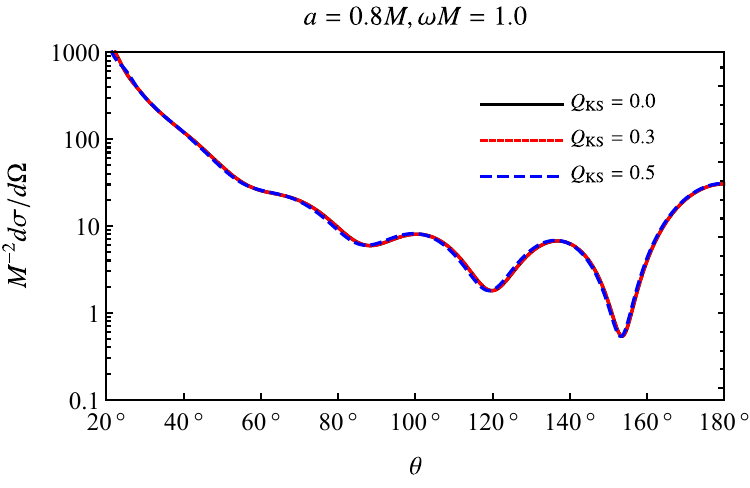}
	\includegraphics[width = 8cm]{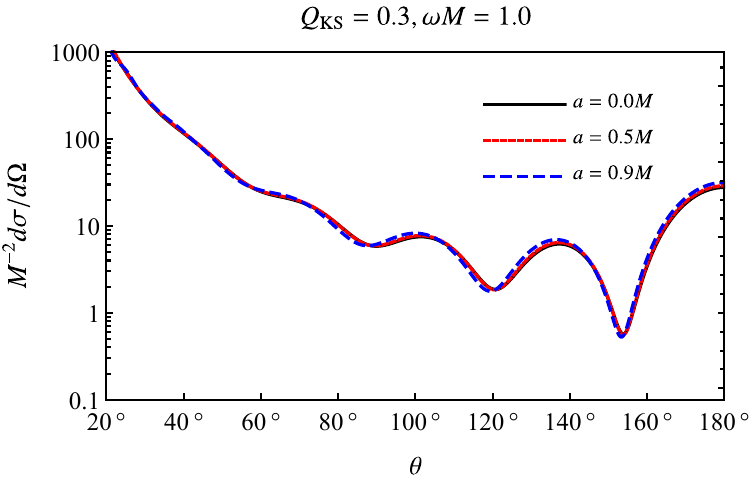}\\
	\includegraphics[width=8cm]{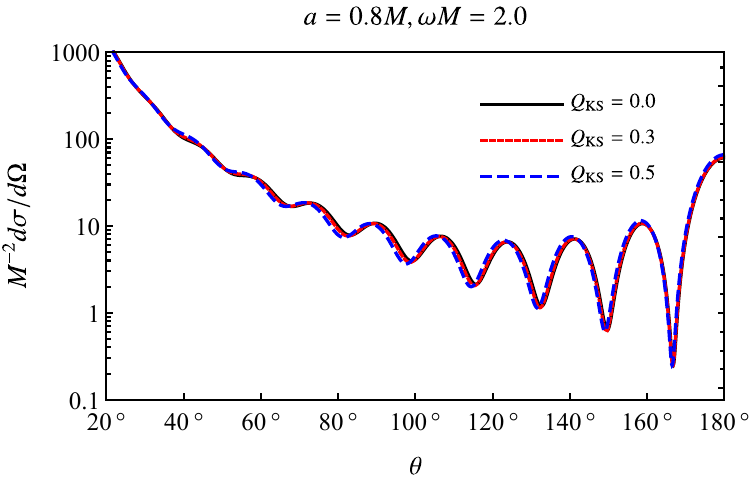}
	\includegraphics[width = 8cm]{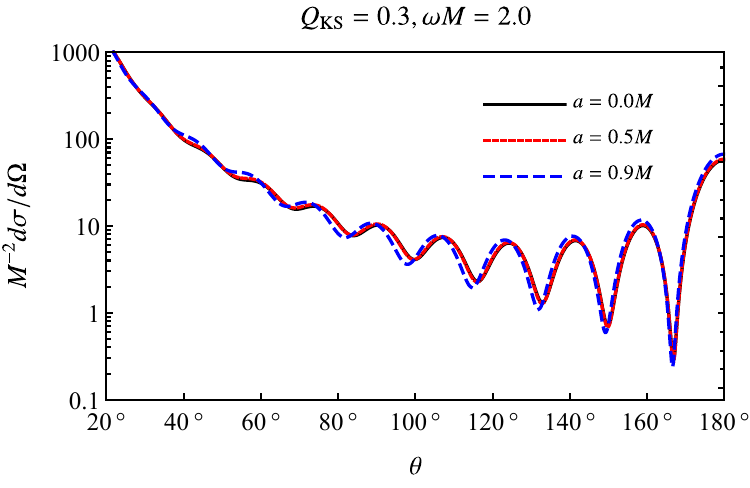}\\
	\includegraphics[width=8cm]{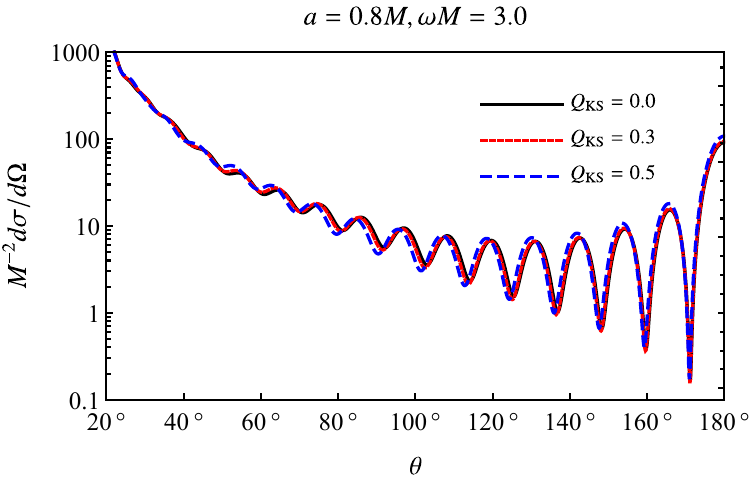}
	\includegraphics[width = 8cm]{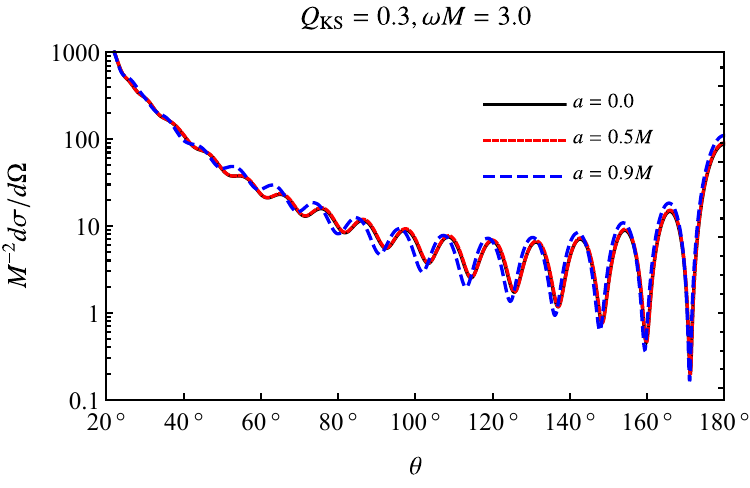}
	\caption{\label{SCS} KS differential scattering cross section with fixed values of $a/M$ (left) and normalized electric charge $\Q$ (right). We exhibit results for three values of frequency: $M\omega=1.0$ (top), $M\omega=2.0$ (middle) and $M\omega = 3.0$ (bottom).}
	\end{figure*}
	
	\begin{figure*}
	\includegraphics[width=8cm]{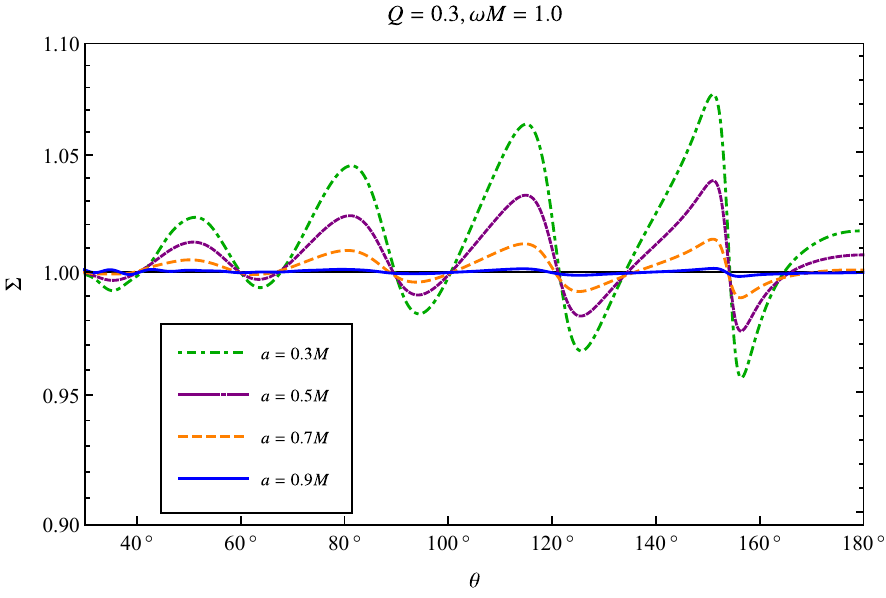}
	\includegraphics[width = 8cm]{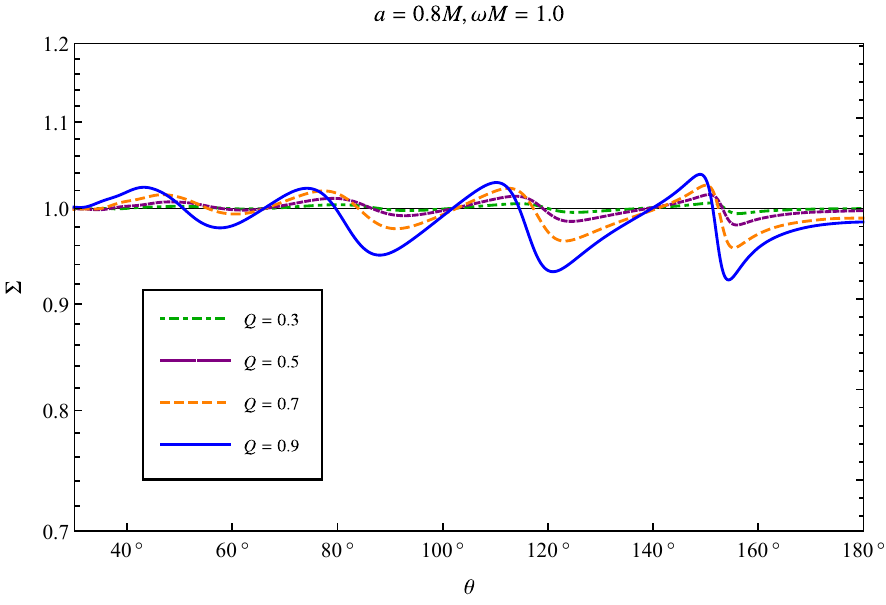}\\
	\includegraphics[width=8cm]{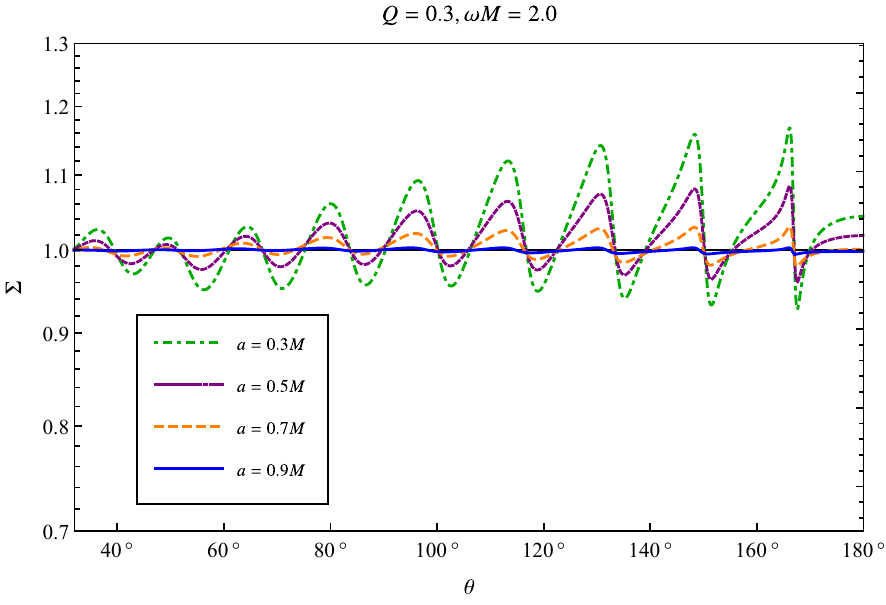}
	\includegraphics[width = 8cm]{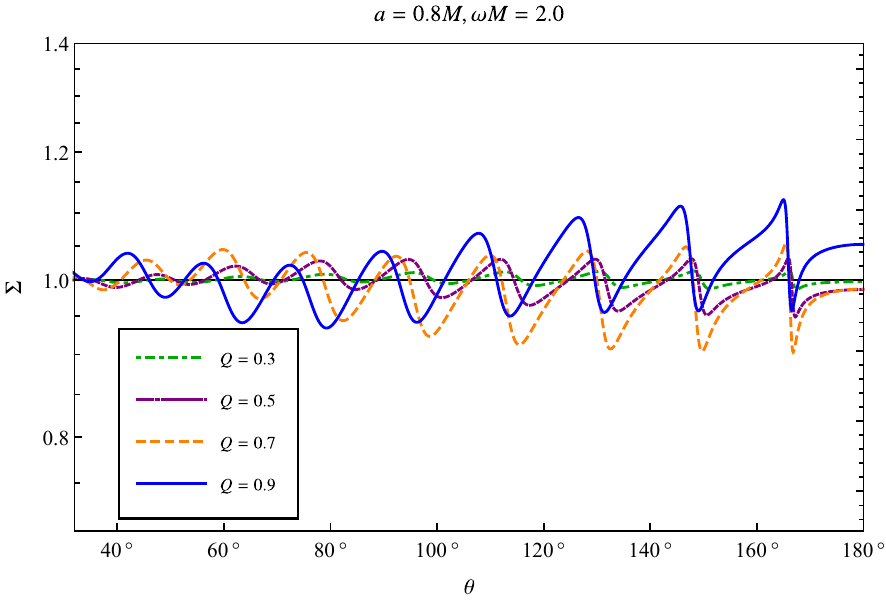}\\
	\includegraphics[width=8cm]{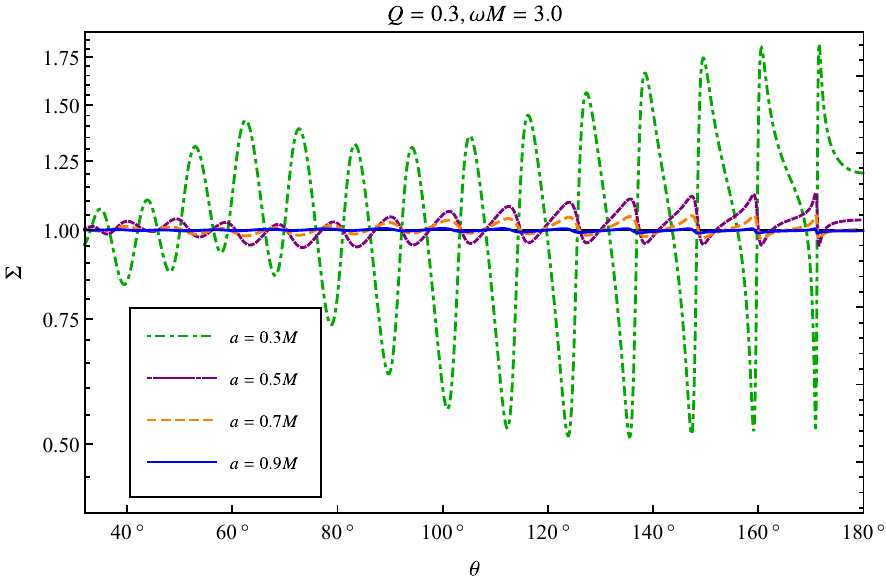}
	\includegraphics[width = 8cm]{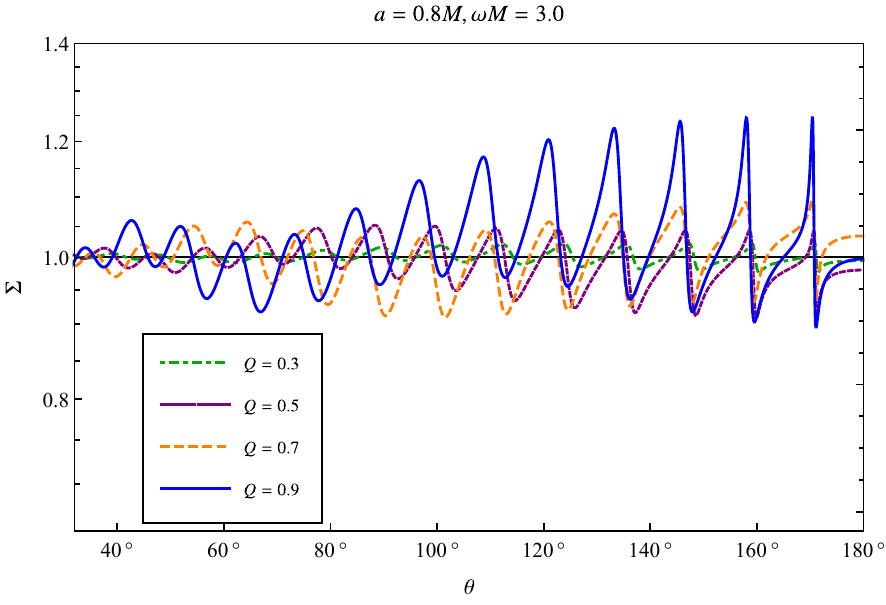}\\
	\caption{\label{comp} Ratio of the scattering cross section of KNBH and KSBH with $\omega M$ = 1.0 (top), $\omega M$ = 2.0 (middle) and $\omega M$ = 3.0 (bottom). The left column shows the results for different values of angular momentum $a/M$ while the right column shows the outcomes for different values of normalized electric charge $\Q$.}
	\end{figure*}
	
In this Section we show a selection of our results for the scattering cross section of the KSBH, in the Einstein frame, for the on-axis incidence of a scalar wave. To obtain our results, we solve Eq.~\eqref{RWeq} numerically using the procedures discussed in Refs.~\cite{Leite_etal:PRD2019,Leite_etal:2019}.  
It is well-known in the literature that the scattering amplitude given in Eq.~\eqref{scatt_amplitude} is poorly convergent. To deal with this shortcoming, we used a series reduction technique described in Refs.~\cite{Leite_etal:2019,Stratton_etal:2020}.

In Fig.~\ref{SCS}, we exhibit the massless and chargeless scalar scattering cross section of KSBH for different values of the normalized electric charge $\Q$ (left) and distinct values of the angular momentum $a/M$ (right). 
We present a selection of our results for three different choices of frequency: $M\omega=1.0$ (top), $M\omega=2.0$ (middle) and $M\omega=3.0$ (bottom). One can note that with the increase of the parameters of the BH, the interference fringes become wider. Also, it is possible to observe a high peak at $180^\circ$, identified as the glory peak. The spectrum for small scattering angles does not change much with the change in the value of the electric charge when compared to the case of an uncharged black hole~\cite{Crispino_etal:2009}. The most relevant changes occur in the regime of large scattering angles. 

	In Fig.~\ref{comp}, we compare the scattering cross section of KNBHs and KSBHs.	We exhibit results for the following values of wave frequency: $M\omega=1.0$ (top), $M\omega=2.0$ (middle) and $M\omega=3.0$ (bottom). Our results show that the difference between the cross sections of corresponding KS and KN BHs is modest. We also notice an oscillation that intensifies as the frequency of the wave increases. This oscillation changes with the BH parameters. 
	
Furthermore, we notice that for low values of electric charge and high values of angular momentum, the ratio between the scalar scattering cross section of the BHs tends to unity, regardless of the wave frequency. This effect can be seen mainly on the left side of Fig.~\ref{comp}, specifically in the case of $a/M=0.9$.	
	
For small values of electric charge, the difference between the scattering cross section of KSBHs and KNBHs is tiny, as can be seen in the right column of Fig.~\ref{comp}. This is consistent with the fact that both KS and KN BHs tend to the Kerr solution as $\Q\rightarrow 0$. The larger differences between the KSBH and KNBH results for lower values of the rotations parameter, seen in the left column of Fig.~\ref{comp}, can be understood due to the fact that, in the regime of staticity ($a\rightarrow 0$), KS and KN BHs tend to different solutions.

\section{Final remarks}\label{sec:remarks}
In recent years, several works have been dedicated to the KS solution. However, to the best of our knowledge, the phenomenon of scattering of fields by KSBH has not yet been studied. In this work, we aimed to fill this gap.

We have investigated the scattering of a massless scalar wave by a charged and rotating BH of heterotic string theory. We computed numerically the scattering cross section of KSBHs for the case of on-axis incidence. Our results show that the enhancement of either the angular momentum or the electric charge increases the widths and amplitudes of the oscillation of the scattering cross section. 
  
We have also compared the KSBH results with the corresponding KNBH ones, showing that the ratio between the scattering cross sections presents an oscillatory behavior with amplitude that increases as (i) the value of the BH electric charge is increased and (ii) the BH angular momentum is decreased.
 
Our results show that KSBHs and KNBHs have very similar scattering cross sections for massless scalar waves. 
 Our findings are complementary to other outcomes in the literature \cite{Xavier_etal:2020,Xavier_etal:2021, Huang_Zhang:2020}~to reinforce the conclusion that it would be challenging to distinguish this stringy black hole from an equivalent BH described by Einstein's theory.

\begin{acknowledgments}
S.~X., C.~B. and L.~C. would like to thank the University of Aveiro, in Portugal, for the kind hospitality during the completion of this work. We acknowledge Funda\c{c}\~ao Amaz\^onia de Amparo a Estudos e Pesquisas (FAPESPA),  Conselho Nacional de Desenvolvimento Cient\'ifico e Tecnol\'ogico (CNPq) and Coordena\c{c}\~ao de Aperfei\c{c}oamento de Pessoal de N\'{\i}vel Superior (Capes) - Finance Code 001, in Brazil, for partial financial support. This work has further been supported by the European Union's Horizon 2020 research and innovation (RISE) programme H2020-MSCA-RISE-2017 Grant No. FunFiCO-777740 and by the European Horizon Europe staff exchange (SE) programme HORIZON-MSCA-2021-SE-01 Grant No. NewFunFiCO-101086251.
\end{acknowledgments}

	{}
\end{document}